\documentclass[twocolumn,showpacs,preprintnumbers,prb]{revtex4}

\usepackage{graphicx}
\usepackage{graphics}
\usepackage{dcolumn}
\usepackage{bm}

\begin{document}
\preprint{AP report no. 03-55}

\title{Scanning tunneling microscopy and spectroscopy of
sodium-chloride overlayers on the stepped Cu(311) surface: Experimental and
theoretical study}

\author{F.~E.~Olsson and M.~Persson}

\address{
Dept. of Applied Physics, Chalmers/G\"oteborg University,
S-41296 G\"oteborg, Sweden.
}

\author{J.~Repp and G.~Meyer}

\address{IBM Research, Zurich Research Laboratory, CH-8803
R\"uschlikon, Switzerland}

\date{\today}


\begin{abstract}
The physical properties of ultrathin NaCl overlayers on the stepped
Cu(311) surface have been characterized using scanning tunneling
microscopy (STM) and spectroscopy, and density functional
calculations.  Simulations of STM images and differential conductance
spectrum were based on the Tersoff-Hamann approximation for tunneling
with corrections for the modified tunneling barrier at larger voltages
and calculated Kohn-Sham states. Characteristic features observed in
the STM images can be directly related to calculated electronic and
geometric properties of the overlayers. The measured apparent barrier
heights for the mono-, bi-, and trilayers of NaCl and the
corresponding adsorption-induced changes in the work function, as
obtained from the distance dependence of the tunneling current, are
well reproduced by and understood from the calculated results. The
measurements revealed a large reduction of the tunneling conductance in a
wide voltage region, resembling a band gap. However, the simulated spectrum
showed that only the onset at positive sample voltages may be viewed
as a valence band edge, whereas the onset at negative voltages is
caused by the drastic effect of the electric field from the tip on the
tunneling barrier.
\end{abstract}
\maketitle


\section{Introduction}
Thin films of insulating materials on metal surfaces are of direct
technological interest in microelectronics, catalysis and as a
corrosive protection. They are also attracting increasing interest for
nanoscience and nanotechnology as potential substrates for atomic and
molecular manipulation by the scanning tunneling microscope. In
particular, this interest stems from the possibility to decrease the
electronic interaction of assembled molecules and adatoms with the
metal support.  The non-vanishing electron density that reaches
through the ultrathin insulating films, thickness below 1\,nm, allows
for these surfaces to be studied by means of
STM~\cite{kn:hebenstreit,kn:schneider,kn:libuda}.  A most important
prerequisite for the realization of this possibility is to be able to
grow stable and atomically thin insulating films on metal surfaces
with a well-characterized and ordered geometric structure. So far the
number of systems studied that fulfills these conditions have been
very limited~\cite{kn:baumer}.  One interesting candidate is provided
by NaCl overlayers on a stepped Cu(311) surface, which we have studied
in detail by scanning tunneling microscopy (STM) ~\cite{kn:repp_311}
and also most recently by density functional
calculations~\cite{kn:olsson_NaCl}.

These experimental and theoretical studies revealed several
interesting aspects of the growth, structure and bonding of these
systems. The initial growth of NaCl on the Cu(311) surface was found to be two
dimensional and commensurate.  The lattice match of the overlayer with
the substrate was suggested in the experimental study to be stabilized
by the incomplete screening of the step Cu atoms resulting in an
electrostatic interaction between these atoms and the Cl
atoms. Density functional calculations corroborated this suggestion
and showed also that the bonding of the overlayer on the surface was further stabilized by the
formation of a weak chemical bond between the step Cu atoms and the Cl
stoms. These calculations revealed also large relaxations of the
monolayers such as a buckling of the layer with a large influence on
the work function. These results call for a detailed scanning
tunneling microscopy study of the topography of and the tunneling
conductance through these overlayers. In particular, the imaging of
insulating overlayers on metal surfaces for biases well below 1 V has not
been studied before.

In this paper, we present atomically resolved topographical images of
NaCl mono-, bi- and tri-layers on Cu(311) obtained by an STM and
discuss them in relation to simulated images based on density
functional calculations and the Tersoff-Hamann (TH)
approximation~\cite{kn:tersoff}. In this approximation, STM images and
scanning tunneling spectra are obtained from the local density of
states (LDOS) at the position of the tip apex.  The conductivity of
the sample are further investigated using scanning tunneling
spectroscopy. The measured $\frac{dI}{dV}$ spectra are compared with
simulated spectra and discussed in relation to the energy dependence
of the calculated LDOS.  This study includes also a calculation and
discussion of the apparent heights and work functions of these layers
obtained from tunneling measurements. The apparent height of an NaCl
overlayer contains information about the ``local'' work function, that
is, the tunneling barrier. The tunneling barriers are obtained from the measured distance
dependence of the tunneling conductance.

STM investigations of insulating overlayers provide important
information about the growth mechanisms and the local electronic
structure of the overlayer.  In earlier STM investigations of
ultrathin insulating overlayers on metal surfaces such as MgO on
Ag(100)~\cite{kn:schneider} and NaCl on Al(111) and Al(100)~\cite{kn:hebenstreit} the
experiments were carried out at relatively high biases of about 1eV to
provide atomic resolution of the overlayer. Atomically resolved STM
images have also been observed for NaCl layers on Ge(100)
\cite{kn:umbach}, where only one type of atoms was
found to be imaged as a protrusion~\cite{kn:hebenstreit}.
 A combined STM and density
functional study of NaCl layers on an Al(111) and Al(100) surface revealed
that only the Cl atoms were imaged as protrusions.  In this work, the
atomic contrast as well as the apparent height of the NaCl layers was
discussed by comparing the experimental images with calculations of
LDOS within the TH approximation.  MgO layers on an Ag(100) surface were studied using
measured $\frac{dI}{dV}$ spectroscopy and the band gap of the bulk MgO
electronic structure was found to be developed within the first three
monolayers \cite{kn:schneider}.  This finding was corroborated by
density functional calculatrions of the energy dependence of the LDOS
within the MgO layers.

The paper is organized as follows. In section~\ref{sec:EXP} our experimental STM and $\frac{dI}{dV}$
results for mono-, bi- and tri-layers of NaCl on Cu(311) are presented. In
section~\ref{sec:theory} we present the geometric and electronic structure and simulated STM images and
$\frac{dI}{dV}$ spectra for the same systems. The experimental and theoretical results are compared and
discussed in Section \ref{sec:DISC}. Finally, section \ref{sec:CONCL} concludes the paper.

\section{Experimental methods and results} \label{sec:EXP}

\begin{figure}
\centering
\includegraphics[width=7cm]{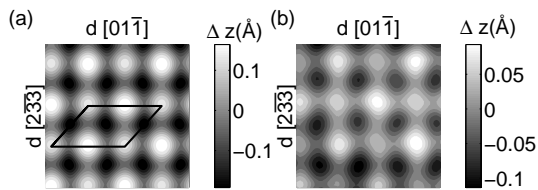}
\caption{Experimental STM images and calculated LDOS images of a
NaCl monolayer in the p(3x2)-II structure. (a) Calculated LDOS
image at an average distance of 7.9 \AA\ from the Cu(311) surface. The surface unit cell used in the
calculation, see Fig. \ref{fig:pos} (right), is indicated. (b)
Experimental STM image with tunneling parameters 210pA and -24meV. In
both (a) and (b) the presented area is 15x15\AA$^2$.
\label{fig:LDOS_c2x2}}
\end{figure}
\begin{figure}
\centering
\includegraphics[width=7cm]{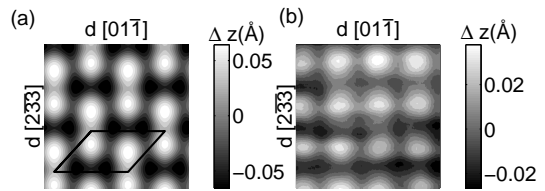}
\caption{Experimental STM images and calculated LDOS images of a
NaCl bilayer on Cu(311). (a) Calculated LDOS image at an
average distance of 9.0\AA\ from the Cu(311) surface. The surface unit cell used in the
calculation, see Fig. \ref{fig:pos} (left), is indicated. (b) Experimental
STM image with tunneling parameters 360pA and -100meV. In both (a) and
(b) the presented area is 15x15\AA$^2$.
\label{fig:LDOS_2}
}
\end{figure}

Our experiments were carried out with a low temperature scanning
tunneling microscope~\cite{kn:meyerstm96}, operated at
13K. We used a chemomechanically polished Cu(311) single crystal,
which was cleaned by Ne$^+$ sputtering and annealing at 750K.  NaCl
was evaporated thermally and the deposition and annealing temperature
was varied in the range of 400K-570K.  In this temperature range Cu
surface atoms are mobile. As an STM tip we used an electrochemically
etched tungsten wire. Bias voltages refer to the sample voltage with
respect to the tip. The STM images were taken at 13K.  The STM images
for mono- and bi-layers that are presented in
Figs. \ref{fig:LDOS_c2x2}(b)-\ref{fig:LDOS_2}(b) and bare Cu(311) were acquired at
relatively low voltages of $\mid V \mid \leq 100$mV, being well within
the band gap of bulk NaCl. For the bare Cu(311) and the monolayer
structures our experimental images were obtained under identical
tunneling conditions, that is, both applied current and voltage, and
tip were identical. Since the growth of the first layer was always
completed before the formation of the second layer we could not be
sure that the tip structure was identical in the imaging of the bare
and the bilayer covered surfaces as for the bare and monolayer-covered
surfaces. However, we could simultaneously record STM images of
islands with one to three layers on top of the monolayer. Up to four
layers of NaCl could be imaged with atomic resolution.

\begin{figure}
\centering
\includegraphics[width=7cm]{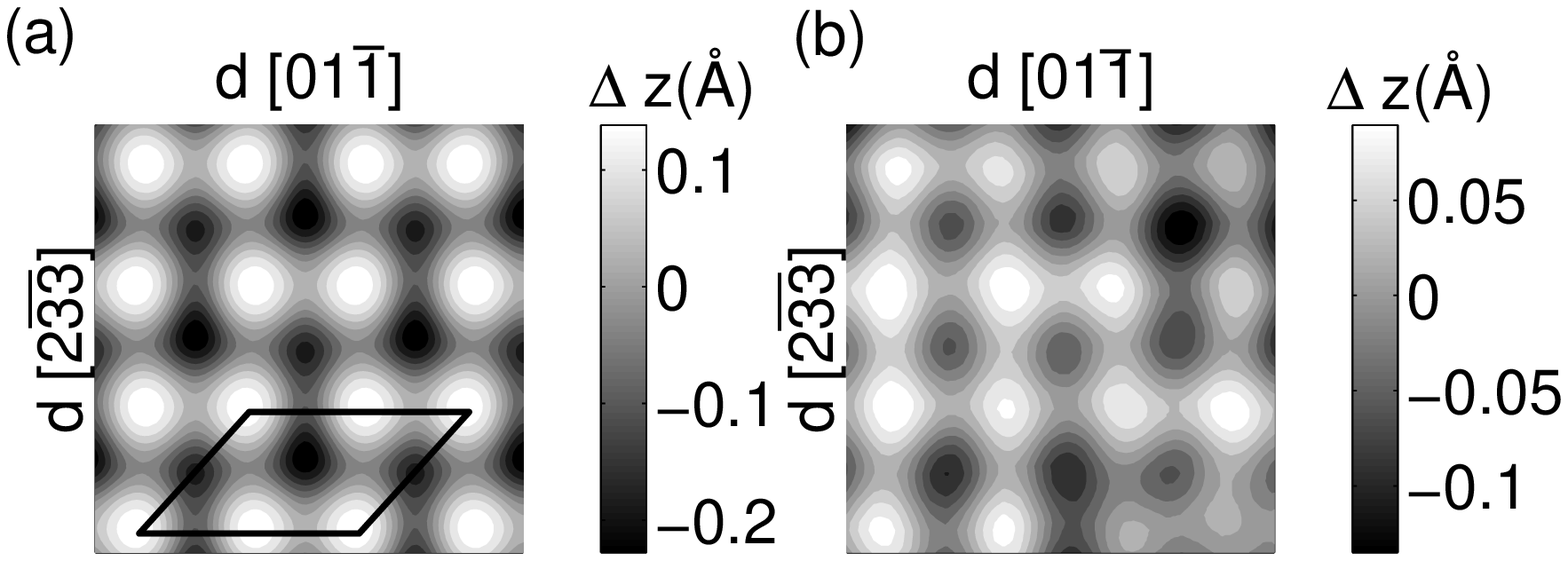}
\caption{Experimental STM images and calculated LDOS images of a NaCl
monolayer in the p(3x2)-I structure. (a) Calculated LDOS image at an
average distance from the Cu(311) surface of 7.9 \AA. The surface unit cell used in the
calculation, see Fig. \ref{fig:pos} (left), is indicated. (b) Experimental
STM image with tunneling parameters 210pA and -24meV. In both (a) and
(b) the presented area is 15x15\AA$^2$.
\label{fig:LDOS_p1x1}}
\end{figure}

In our experiments we observe two different structures for the NaCl
monolayer, shown in Figs. \ref{fig:LDOS_c2x2}(b) and
\ref{fig:LDOS_p1x1}(b).  As argued in our earlier study, we attribute
these images to be associated with the structures presented in
Fig. \ref{fig:pos}. Here we follow the notation of Ref.
\cite{kn:olsson_NaCl} and designate these structures as p(3x2)-II and
p(3x2)-I, respectively~\cite{kn:note1}.
In the STM images of these structures, we identify one characteristic
feature for each structure. In the STM image of the p(3x2)-II
structure every second protrusion appear slightly brighter.  For the
p(3x2)-I structure we note a slight tendency for the protrusions to
group into pairs, even though, this effect is small and is just at the
limit of the present resolution. In contrast to the monolayer,
atomically resolved STM images of the bi- and trilayers
(Fig. \ref{fig:LDOS_2}(b)) did not show any differences between the
domains corresponding to p(3x2)-I and -II structures, although it
could be concluded that both these domains are present with the help
of characteristic defects, consisting of a missing Cu atoms in the
topmost substrate layer \cite{kn:repp_311}.

The apparent heights, $\Delta z_l$, of the NaCl overlayers ($l=1,2,3$
for the mon-, bi-, and tri-layer) are presented in
Table~\ref{tab:workfunction}. $\Delta z_l$ is defined as the vertical
change in tip position when a step edge is scanned. We find that
$\Delta z_l$ decreases with {\em l} and is positive for all {\em
l}. As also shown in Table \ref{tab:workfunction}, the geometric
heights, $\Delta z^{\rm geom}_l$, of the NaCl overlayers are about
1\AA\ larger than $\Delta z_l$.  This implies that the tip-surface distance
decrease with {\em l}.
\begin{table}
\caption{Calculated work function $\Phi_{\rm theory}$ and
experimentally measured mean barrier height $\overline{\Phi}_{\rm
exp.}$ of NaCl layers on Cu(311). The values in brackets indicate the
relative changes with respect to the clean copper surface.  $\Delta
z_l$ are the apparent heights obtained from LDOS calculations (theory)
and STM experiments (exp) and $\Delta z_l^{geom.}$ are the calculated
geometric height. {\em l}=1,2 and 3 for mono-, bi- and trilayers,
respectively. The result for the NaCl monolayer is for the p(3x2)-II
structure.
\label{tab:workfunction}
}
\begin{tabular}{lcccc}
                                &Cu(311)& mono    &  bi     & tri \\
\hline
$\Phi_{\rm theory}$ (eV)         & 4.31  & 3.53 (-18\%)& 2.85 (-34\%)& 2.83 (-34\%)\\
$\overline{\Phi}_{\rm exp.}$(eV) & 3.81 & 3.23 (-15\%)& 2.82 (-26\%) & 2.82 (-26\%) \\
\hline
$\Delta z_{l}$ (exp) (\AA)  & &1.6      & 1.4     & 1.3  \\
$\Delta z_{l}$ (theory) (\AA)& &1.8      & 1.1     & -    \\
$\Delta z_l^{\rm geom.}$ (\AA) & &2.5      & 2.8     & 2.8    \\
\end{tabular}
\end{table}

To quantify work function changes upon adsorption of NaCl layers, we
have carried out measurements of the distance dependence of the
tunneling current $I$\cite{kn:binnig82}.  The feed-back loop was
switched off and a voltage ramp was applied to the $z$-piezo drive
while measuring the current $I$ as a function of tip-sample distance
$z$. The resulting sweep in $z$ measured 2.7 {\AA} in height and had a
duration of about one second. The resulting current curve followed
closely an exponential dependence, $I(z) \propto \exp(-2\kappa z)$
from which an apparent barrier height $\overline{\Phi}$ was defined
from the decay constant $\kappa$ as $\overline{\Phi} =
\hbar^2\kappa^2/(2m_e)$. The relationship between $\overline{\Phi}$
and the work function $\Phi$ of the sample is not straightforward and
has been discussed at length in the
literature~\cite{kn:lang88,kn:olesen96}. The systematic deviation of
$\overline{\Phi}$ from $\Phi$ decreases with increasing tip-surface
distance, corresponding to a larger vacuum region between the tip and
the sample. So this deviation was minimized by recording spectra at a
lowest possible current, ranging down to 10pA, so as to maximize the
tip-surface distance.  Although $\Phi$ can not be determined directly,
the adsorption-induced relative changes in $\overline{\Phi}$ are
expected to correspond to the relative changes in $\Phi$.  The
experimental uncertainty in determining $\overline{\Phi}$ from
$\kappa$ is set by the vertical length scale calibration, being a few
percent. Using this procedure, we find that the measured work function
difference between a NaCl monolayer and the clean surface agrees
well with the work function difference measured by an
atomic force microscopy tip~\cite{kn:bennewitz}.

The apparent barrier height $\overline{\Phi}$ for the bare surface and
NaCl overlayers deduced from the measured the $I(z)$ are shown in
Table~\ref{tab:workfunction}. These results show that there is a
substantial decrease in $\overline{\Phi}$ upon adsorption of the first
layer and an even larger decrease for the second. No further changes
of $\overline{\Phi}$ are observed for the thicker layers. This result
indicates that the nature of the monolayer differ from the interface
layer in the multilayers.

\begin{figure}
\centering
\includegraphics[width=7cm]{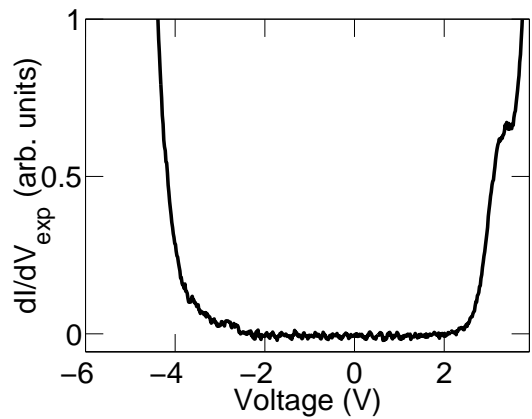}
\caption{Experimental dI/dV spectrum for a monolayer NaCl on Cu(311). The
tip-sample distance corresponds to a current setpoint of $I=0.3$ nA at $V=3.8$ V. This distance is by
about 6 \AA\ larger than the distance for the imaging parameter set used in the Figs.
\protect{\ref{fig:LDOS_c2x2}} and \protect{\ref{fig:LDOS_p1x1}}.
\label{fig:STS_exp}
}
\end{figure}

The insulating properties of the monolayer have been investigated
using dI/dV spectroscopy, shown in Fig.~\ref{fig:STS_exp} for a
monolayer. When acquiring the spectra the tip was retracted
about 6\AA. The spectra is not sensitive to the tip position relative
to the surface.  The poor conductivity of the
monolayer is manifested by a very small tunneling current between -4
and +3V and exponential off and onsets of $\frac{dI}{dV}$ at
these voltages. The other characteristic feature in the spectra is the
shoulder at about +3.2V.

\section{Theoretical methods and results}\label{sec:theory}

\begin{table}
\caption{Calculated interface energies and geometrical parameters of a
NaCl monolayers, bilayer and trilayer on the (311) surface of Cu.  The
interface layer for the bilayer and trilayer correspond the p(3x2)-I
structure. The adsorption energy, E$_a$, is given in eV per NaCl pair
of the interface and is defined as
$E_a=(E_{lNaCl/Cu}-E_{nNaCl}-E_{(l-n)NaCl/Cu}$)/m, where l is the
number of NaCl layers in the overlayer, m is the number of NaCl in the
interface and n=1, n=1,2 and n=1,2,3, for monolayers, bilayers and
trilayers, respectively. $\Delta b = z_{Cl}-z_{Na}$, where z$_{Cl}$
and z$_{Na}$ are the coordinates of Na and Cl atoms perpendicular to
the surface. $z_{Cl}$ is the distance between the Cl atom and the Cu
surface layer.  d$_{Cl-Cl}$ is a Cl--Cl distance along a row in the
layer. The distance between Cl atoms on two neighbouring Cu-rows is
4.26 \AA.  The superscripts $(t)$ and $(b)$ for the p(3x2)-I structure
corresponds to Cl atoms in top and bridge sites relative to the top
Cu(311) layer, respectively. The superscript $(t')$ and $(b')$ for the
p(3x2)-II structures corresponds to Na atoms in top and bridge sites
relative to the second Cu(311) layer, respectively. See Fig. 1 in
Ref. \protect{\cite{kn:olsson_NaCl}}.
\label{tab:positions}
}
\begin{tabular}{lcccc}
         & E$_{a}$ (eV) & $\Delta$ b (\AA)  & $z_{Cl}$ (\AA) &  d$_{Cl-Cl}$\\
\hline
\multicolumn{4}{l}{monolayer } \\
\hline
p(3x2)-I   & 0.42  & 0.31$^{(t')}$, 0.35$^{(b')}$     & 2.54   & 3.78, 3.94\\
p(3x2)-II  & 0.41  & 0.36$^{(t)}$, 0.29$^{(b)}$       & 2.56$^{(t)}$ 2.49$^{(b)}$ & 3.86, 3.86\\
\hline
\multicolumn{4}{l}{bilayer } \\
\hline
interface & 0.35  &0.13$^{(t')}$, 0.11$^{(b')}$ & 2.50$^{(t)}$   & 3.83, 3.89 \\
top      & 0.22   & $<$ 0.02                    & 5.29           & 3.86, 3.86  \\
\hline
\multicolumn{4}{l}{trilayer } \\
\hline
interface & 0.34 & 0.15$^{(t')}$, 0.13$^{(b')}$  & 2.51 & 3.83, 3.89 \\
middle   & 0.25  & -0.09               & 5.25          & 3.86, 3.86 \\
top      & 0.31   & 0.08               & 8.14          & 3.86, 3.86 \\
\end{tabular}
\end{table}

\begin{figure}
\centering
\includegraphics[width=7cm]{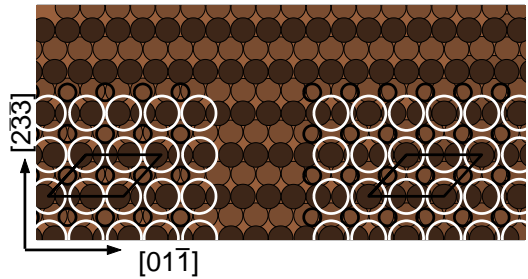}
\caption{Ordered structures of a NaCl monolayer on Cu(311): (left)
p(3x2)-I and (right) p(3x2)-II. The dark and light filled circles
represent Cu atoms of the first and second layer, respectively, and
the large white and small black open circles represent Cl and Na
atoms, respectively.  The unit cells used in the calculations are indicated.
In the p(3x2)-I structure all Cl atom sites are
equivalent, whereas there are two inequivalent Na atom sites, top (t')
and bridge (b'), with respect to the second Cu layer. In the p(3x2)-II
all Na atom sites are equivalent, whereas there are two inequivalent
Cl atom sites, top (t) and bridge (b) sites with respect to the first
Cu layer, see also Fig. 1 in Ref. \protect{\cite{kn:olsson_NaCl}}.
\label{fig:pos}
}
\end{figure}

The results obtained from the STM measurements for the bare surface
and the NaCl overlayers such as constant current images, work
functions and apparent heights have been analyzed by extending our
recent density functional calculations of the electronic and geometric
structures of these system~\cite{kn:olsson_NaCl}, including now also
the trilayer, to tunneling.  These calculations were based on a plane
wave basis set and the projector augmented wave (PAW)
method~\cite{kn:kresse,kn:blochl} as implemented in the {\tt
VASP}~\cite{kn:kresse_VASP} code. A generalized gradient
approximation~\cite{kn:pw91} was employed for the exchange-correlation
potential. The system was represented by a slab in a super cell
geometry~\cite{kn:note2} and the equilibrium geometry was
obtained by a structural optimization~\cite{kn:note3}. We considered the two p(3x2)-I and
p(3x2)-II structures for the NaCl monolayer, shown in
Fig.~\ref{fig:pos} and the structures for the bilayer and trilayer
that were formed by adding subsequent NaCl layers on the p(3x2)-I
structure of the monolayer.  In Tables~\ref{tab:workfunction} and
~\ref{tab:positions}, we show the calculated work functions and key
bonding parameters for the NaCl layers. For easy reference these
tables include also the parameters for monolayer and the bilayer,
previously presented in Ref.~\cite{kn:olsson_NaCl}. The calculated
partial $s$, $p$, and $d$ density of states of the Cl atoms in
trilayer are shown in Fig. \ref{fig:PDOS}. As expected, the width
of the Cl peak is largest for the interface layer
(Fig. \ref{fig:PDOS}(c)) and smallest for the outermost layer
(Fig. \ref{fig:PDOS}(a)), due to larger overlap with the Cu states for
the interface layer.

The STM images and the apparent heights of the overlayers were
simulated using the Tersoff and Hamann (TH)
approximation~\cite{kn:tersoff} for the tunneling current. In this
approximation the tunneling current for low bias is proportional to
the LDOS, $\rho({\bf r_0},\epsilon_F)$, of the surface at the position
of the tip apex, {\bf r}$_0$. The LDOS was calculated from the Kohn-Sham
wavefunctions, see Ref. \protect\cite{kn:olsson} for details. The topographical STM
images were obtained from the topography of the contours of constant
$\rho({\bf r_0},\epsilon_F)$, which henceforth will be referred to as
LDOS images.

We have also simulated differential conductance, $\frac{dI}{dV}$ spectra, using the LDOS outside the
surface. For these simulations it is necessary to consider the
effects on the tunneling caused by the applied electric field from the tip. Following
Lang \cite{kn:langdIdV}, these effects are partly accounted for by calculating
$\frac{dI}{dV}$ within the average tunneling barrier approximation as
\begin{widetext}
\begin{eqnarray}
\frac{dI}{dV} & \propto & \frac{d}{dV}
(\int_{\epsilon_F}^{\epsilon_F+eV}\rho_S(\epsilon)T(\epsilon,V)d\epsilon)
\label{eq:dIdV} \\
& = & e\rho_S(\epsilon_F+eV)T(\epsilon_F+eV,V) +
\int_{\epsilon_F}^{\epsilon_F+eV}\frac{d}{dV}(\rho_S(\epsilon)T(\epsilon,V))d\epsilon \,,
\label{eq:dIdV_lang}
\end{eqnarray}
\end{widetext}
where $T(\epsilon,V)$ is the transmission coefficient through the
average barrier for an electron with energy $\epsilon$ and is given
by,
\begin{equation}
T(\epsilon,V) =
\exp{(-2s\sqrt{2m(\Phi_{av}+\epsilon_F-\epsilon+eV/2)}/\hbar)}
\label{eq:Tdef}
\end{equation}
Here $\Phi_{av}= (\Phi_T+\Phi_S)/2$ where $\Phi_T$ and $\Phi_S$ is the
tip and sample work function, respectively, $\rho_S(\epsilon)$ is the
LDOS at a distance $z_s$ just outside the surface layer defining the
boundary of the vacuum region and $s$ is the distance between $z_s$
and the tip apex. The first term in
Eq.~(\ref{eq:dIdV_lang}) is the LDOS at the tip-apex taking into
account electric field in the tunneling junction. In the limit of
vanishing voltage only this term survives and the TH approximation is
recovered. Note that this result for $\frac{dI}{dV}$ is well defined
for $|V| <  2\Phi_{av}$ but the average barrier approximation breaks
down for voltages approaching the field-emission regime, that is, for
$V > \Phi_S/e$ or $V < -\Phi_T/e$.

\begin{figure}
\includegraphics[width=7cm]{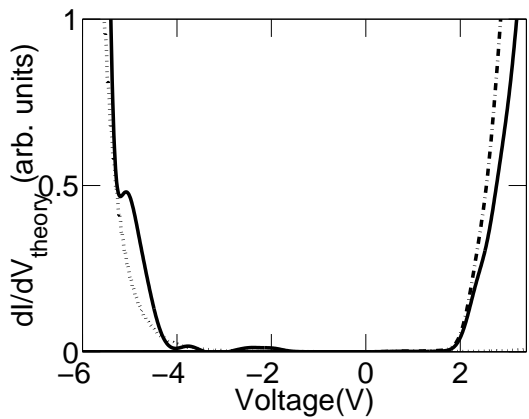}
\caption{Calculated $\frac{dI}{dV}$ spectra for a monolayer NaCl in
the p(3x2)-I structure on Cu(311). The solid and dotted lines are calculated from
Eq.~(\protect\ref{eq:dIdV_lang}) using the full energy dependence of $\rho_S(\epsilon)$ and
a constant $\rho_S(\epsilon)$, $\rho_S(\epsilon)=\rho_S(\epsilon_F)$, respectively.
The dashed-dotted line is
the result obtained using only the local density of states
contribution to $\frac{dI}{dV}$ (first term in
Eq.~(\protect\ref{eq:dIdV_lang})). In the calculations we used $z_s$ = 2 {\AA}, $s$ = 10 {\AA} and
$\Phi_{av}$ = 3.5 eV.
\label{fig:STS_th}
}
\end{figure}
In Fig.~\ref{fig:STS_th}, we show the simulated $\frac{dI}{dV}$
spectra for a monolayer of NaCl in the p(3x2)-II structure. The
behavior of $\frac{dI}{dV}$ in the voltage region -3 to 3 V is well
understood from the behaviour of the LDOS. In particular, the rapid
increase of $\frac{dI}{dV}$ around $V$ = 2 V is an LDOS effect and may
be viewed as conduction band edge.  In contrast, the rapid decrease of
$\frac{dI}{dV}$ around $V$ = -4.5 V is not an LDOS effect associated
with any valence band edge. Note that the large contribution of
electron states of $p$ or $s$ character around the Cl atoms to the
partial DOS in Fig.~\ref{fig:PDOS} in the corresponding energy region does
not show up in the LDOS. As illustrated by the result for
$\frac{dI}{dV}$ in Fig.~\ref{fig:STS_th}, using an energy-independent
$\rho_S(\epsilon) = \rho_S(\epsilon_F)$, the rapid decrease comes from
the second term in Eq.~(\ref{eq:dIdV_lang}),
which describes the drastic effect of the voltage dependent barrier on
the tunneling and the larger probability for tunneling of electrons
with energies closer to the vacuum level at larger negative
voltages. In the calculation we have used the same work function for
the tip and the sample so we expect even a more dramatic onset of
$\frac{dI}{dV}$ for $V <$ -3.5 V corresponding to the field-emission
regime.
\begin{figure}
\centering
\includegraphics[width=7cm]{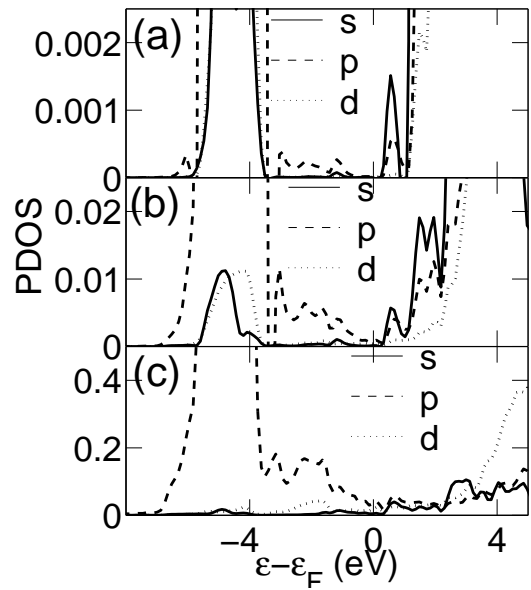}
\caption{Partial $s$, $p$ and $d$ density of states around Cl atoms in
a NaCl trilayer on Cu(311). The Cl atom is in (a) the third
(outermost) NaCl layer, (b) the second layer, and (c) the first
(interface) layer. Note that the scale of the presented PDOS on the
three Cl atoms differ by several orders of magnitude.
\label{fig:PDOS}
}
\end{figure}

\section{Discussion} \label{sec:DISC}

We begin by comparing and discussing the measured apparent
barrier heights and calculated work functions for the
p(3x2)-I monolayer, the bilayer and the trilayer. As shown in
Table~\ref{tab:workfunction}, the calculated changes in work functions
reproduce nicely the measured changes for the bare surface and its
change upon adsorption of the overlayers. Note that the measured
apparent barrier height and the calculated work function can not be
compared directly.  From the calculations we are able to understand
the reduction of $\Phi$ upon adsorption and its layer dependence in
terms of bonding and buckling of the monolayer. This explanation was
discussed in detail in our previous study~\cite{kn:olsson_NaCl} and
will only be summarized here.

In general, the surface contribution to $\Phi$ is determined by the
surface dipole layer, which have contributions both from electronic
and ion-core rearrangements upon adsorption. The reduction of $\Phi$
upon adsorption was argued to be caused by the formation of a weak
chemical bond between the Cl atoms and the metal substrate and the
screening of the Na$^+$ ion by the metal electrons. The substantially
smaller reduction of $\Phi$ for the monolayer than for the bilayer and
the trilayer is caused by the large buckling of the monolayer with
a large inward relaxation of the Na$^+$ ions relative to the Cl$^-$
ions, resulting in a dipole layer that increases the work
function. This buckling is substantially reduced for the bilayer and
the trilayer and the work function reductions for these layers are governed mostly by the
electronic
rearrangements, which are limited to the interface region. The
measured $\Phi$ is thus consistent with a buckling of the monolayer and an
interface bonding that involves electron rearrangements between the
overlayer and the substrate.

Next we turn to a comparison and a discussion of observed and
simulated STM images for the bare surface and the overlayers. These
images contain information about the surface electronic and geometric
structure through the tails of the wave functions in the vacuum
region. Since the TH approximation for the tunneling is not based on
any detailed model for the tip $s$ density of states, we are not able
to determine the distance, $z_0$, between the tip apex and the surface
Cu layer from the tunneling conductance. However, the bare surface and
the two monolayers were imaged under identical tunneling conditions so
we can use the image for the bare surface to calibrate the value of
the $\rho({\bf r_0},\epsilon_F)$ in the calculation of the LDOS images
and to make a prediction of the images for the monolayers.

For the bare Cu surface, the calculated LDOS image gives the same
tip-surface corrugation across the Cu rows as the STM image at an
average $z_0$ of about 6.1 \AA. No corrugation could be resolved along the
Cu rows, neither in the LDOS image nor in the STM image. In contrast to
the flat single crystal metal surfaces for which the TH approximation
fails to reproduce atomically resolved STM images~\cite{kn:chen_prb},
the measured tip-surface corrugation can be
reproduced by a realistic value for $z_0$, for which the TH approximation
should be applicable. For the flat metal surfaces, the imaging mechanism
has been argued to involve a direct tip-surface
interaction~\cite{kn:hofer_prl}.

In Figs.~\ref{fig:LDOS_c2x2} and ~\ref{fig:LDOS_p1x1}, we show the
calculated LDOS images for the investigated monolayers, using the same
value of $\rho({\bf r_0},\epsilon_F)$ as for the bare
surface that reproduced the observed surface corrugation. The
corresponding average value of $z_0$ is 7.9 {\AA}. The resulting LDOS
images are in overall good agreement with the observed STM images
although the corrugation is somewhat larger in the LDOS images than in
the STM images. The LDOS images support our original interpretation of
the observed STM images that the protrusions correspond to Cl
atoms~\cite{kn:repp_311}. The dominant contribution from the states
of the Cl atom to the tunneling is understood from the participation of
the $p$ states in the bond as indicated by the corresponding partial density
of states in Fig.~\ref{fig:PDOS}.

As shown in Fig.~\ref{fig:LDOS_c2x2}, the LDOS image for the p(3x2)-II
monolayer reproduces nicely the experimental observation that every
second Cl is imaged slightly brighter.  In our calculations we
attribute the brighter protrusions to be derived from Cl atoms at
bridge sites. This difference in brightness corresponds to a
difference in $z_0$ of about 0.054 {\AA} and is consistent with larger
adsorption height, $z_{Cl}$, of 0.07 {\AA} (see Table
\ref{tab:positions}) for the Cl atoms at the bridge sites than at the
top sites. However, the hybridization of the $3p$ states of the Cl
atoms with the metal states at the Fermi level is found to be stronger
for the top site than for the bridge site. This electronic effect
counteracts in part the geometric effect on the tip-surface
corrugation.

As shown in Fig.~\ref{fig:LDOS_p1x1}, the tendency of pairing of the
protrusions in the STM image for the p(3x2)-I monolayer is well
reproduced by the LDOS image. This pair formation reflects the
geometric relaxations of the Cl atoms along the Cu rows, resulting in
two different Cl--Cl distances, which differ by 0.16 {\AA} (see Table
\ref{tab:positions}). However, the geometric relaxations contribute
only a minor part to the difference in distance of about 0.9 {\AA}
between protrusions along the rows. So in this case the dominant part
of this difference is an electronic structure effect associated with
the tilting of the Cu--Cl bond and the hybridized $3p$ states of Cl
with respect to the surface normal.

In comparing STM and LDOS images of the bilayer we cannot adopt the
same unambiguous procedure for the choice of average $z_0$ as we used
for the monolayers. The bare surface and the bilayer are never imaged
under the same tunneling conditions because the monolayer is fully
developed before the formation of the bilayer. However, in our
comparison with the STM image we use the same value for $\rho({\bf
r_0},\epsilon_F)$ as above, which corresponds to an average $z_0$ of
9.0 {\AA}. As for the monolayers we find that the protrusions in the
LDOS image for the bilayer shown in Fig. \ref{fig:LDOS_2} correspond
to Cl atoms. As shown in Fig.~\ref{fig:PDOS}, there is still a
contribution from Cl states to the partial DOS around the Fermi
level. This contribution is much smaller than for the Cl atom in the
interface layer but is compensated by the average distance of the tip
from the outermost NaCl layer being only about 3.5 {\AA} for the
bilayer, which is about 2{\AA} shorter than the corresponding distance
for the monolayer with the same value of $\rho({\bf
r_0},\epsilon_F)$. The agreement between the LDOS and STM image with
respect to amplitude of the surface corrugation is about the same as
for the two monolayers but in contrast to the STM image the
protrusions are more diffuse and paired into dimers in the LDOS
image. This pairing is a pure electronic structure effect because the
distances between Cl atoms across the rows are the same even though
the atoms are inequivalent. Note that the distance of about 3.5 {\AA}
between the tip and the outermost NaCl layer is slightly too short to
justify the TH approximation and an enhancement of the contrast from tip-surface interaction
cannot be ruled out~\cite{kn:hofer_prl}. In particular, the tip-surface distance is even
shorter for the tri- and quadrolayers so the observed atomic resaolution might be caused by
the tip-surface interactions being sensitive to the identity of the underlying atom.

The apparent heights $\Delta z_l$ of the single steps on the NaCl
multilayer $l$ provide indirect information about the conductance
through the layers. For example, $\Delta z_l$ is equal to zero or the
geometrical height for a multilayer $l$ of vacuum or Cu,
respectively. Because bulk NaCl is a wide-gap insulator with a
calculated band gap of 5.0 eV~\cite{kn:note4}, the electrons around the
Fermi level have to tunnel through the film so its conductance should
be substantially less than an Cu layer, resulting in a smaller
apparent height than the geometrical height.

As shown in Table~\ref{tab:workfunction}, the calculated $\Delta z_l$
are in good agreement with the experimentally measured $\Delta z_l$
for the mono- and bi- layers and are indeed appreciably smaller than
their geometrical heights. In particular, we find that an adsorbed
multilayer corresponds to a vacuum layer with an height being about
half the geometrical height of the multilayer.  The increasing
difference in geometrical height and apparent heights with the number
of layers makes it increasingly difficult to image the layer.  The
larger $\Delta z_1$ than $\Delta z_l$ for $l>1$ is consistent with the
behavior of the calculated partial $s$, $p$, and $d$ DOS around the Cl
atoms, as shown in Fig.~\ref{fig:PDOS}. The interface layer has the
largest conductivity because of the formation of a weak covalent bond
between a Cl atom in the interface layer and the Cu substrate results
in a non-vanishing density of states in the gap region in the
interface layer.  Whereas for the other layers, this contribution to
the density of states are vanishingly small resulting in a smaller
conductance and smaller apparent heights than for the interface
layer. Note that the variation of the the work functions with the
number of adsorbed layers contribute to $\Delta z_l$ but cannot
explain the variation in $\Delta z_l$ with $l$. For instance, the
decrease of the work function from the monolayer to the bilayer
decreases the difference between $\Delta z_1$ and $\Delta z_2$ because
wave functions for electrons at the Fermi level decreases more slowly
with decreasing work function.

Finally, how the tunneling through a NaCl overlayer is affected by the
presence of a band gap in bulk NaCl is obtained from the measured
$\frac{dI}{dV}$ spectrum. The characteristic features of this spectrum
such as the strong reduction of $\frac{dI}{dV}$ for voltages in the
range from -4.5 to 2 V is reproduced rather well by the simulated
$\frac{dI}{dV}$ spectrum. This drastic reduction suggests a depletion
of the LDOS in the corresponding energy region, which can be
attributed to the absence of propagating states in the overlayer in
this energy region and the formation of a band gap. However, our
discussion in Section~\ref{sec:theory} of the simulated $\frac{dI}{dV}$
spectrum in Figs.~\ref{fig:STS_th} showed that this suggestion is
oversimplified. The onset at around $V$ = 2 V is an LDOS effect
whereas the onset at $V$ = -4.5 V is caused by the voltage-dependent
barrier and field-emission~\cite{kn:note5}.

\section{Conclusions} \label{sec:CONCL}

In conclusion, we have carried out a combined scanning tunneling
microscopy and density functional study of the surface topography, the
apparent heights, work functions and the differential conductance of
mono-, bi-, and tri-layers of NaCl on a stepped Cu(311) surface. The
experimental results are well reproduced by the calculations, which
also provide physical insights and interpretation of the results. The
adsorption-induced changes of electronic and geometric structure of
the NaCl overlayer are found to be significant. The observed large
adsorption-induced changes in the work function are found to be
associated with substantial charge rearrangements upon adsorption. The
Cl atoms were shown to be imaged as protrusions. In the STM images, also the different
positions of the Cl atoms relative to the underlying Cu(311) lattice
in the p(3x2)-I and p(3x2)-II structures of the NaCl monolayer were resolved. The poor
conductivity of the layers
were revealed by the apparent heights of steps being only half of the
geometrical heights. Furthermore, $\frac{dI}{dV}$ spectra revealed a large reduction of the
tunneling conductance in a wide voltage region, resembling a band
gap. However, the simulated spectrum showed that only the onset at positive
sample voltages may be viewed as a conduction band edge, whereas the
onset at negative voltages is caused by the drastic effect of the
electric field from the tip on the tunneling barrier.

\section*{Acknowledgements}

We gratefully acknowledge partial funding by the EU-RTN project
``AMMIST'' and the Deutsche Forschungsgemeinschaft Project No. RI
472/3-2 (JR,GM). FO and MP are grateful for support from the Swedish
Research Council (VR) and the Swedish Foundation for Strategic
Research (SSF) through the materials consortium
``ATOMICS''. Allocations of computer resources through the Swedish
National Allocations Committee (SNAC) and the Consortium of Heavy
Computing (KTB) at Chalmers are also gratefully acknowledged.




\begin{thebibliography}{00}

\bibitem{kn:hebenstreit} W. Hebenstreit, J. Redinger, Z. Horozova, M. Schmid, R. Podlucky
and P. Varga, Surf. Sci {\bf L321} (1999) L321

\bibitem{kn:schneider} S. Schintke, S. Messerli, M. Pivetta, F. Patthey, L. Libioulle, M. Stengel,
A. DeVita and W D Schneider, Phys. Rev. Lett {\bf 87} (2001) 276801

\bibitem{kn:libuda} J. Libuda, F. Winkelmann, M. B\protect{\"a}umer, H. J. Freund, T. Bertams,
N. Neddermeyer and K M\protect{\"u}ller, Surf. Sci., {\bf 318} (1994) 61

\bibitem{kn:baumer} M. B\protect{\"a}umer and H. J. Freund, Progr. Surf. Sci. {\bf 61} (1999)
127

\bibitem{kn:repp_311} J. Repp, S. F\protect{\"o}lsch, G. Meyer and K.-H. Rieder, Phys. Rev.
Lett. {\bf 86} (2001) 252

\bibitem{kn:olsson_NaCl} F. E. Olsson and M Persson, Surf. Sci. {\bf 540} (2003) 172

\bibitem{kn:tersoff} J. Tersoff and D. R. Hamann, Phys. Rev. Lett. {\bf 50} (1983) 1998


\bibitem{kn:umbach} K. Gl\protect{\"o}ckler, M. Sokolowski, A. Soukopp, E. Umbach, Phys. Rev.
B {\bf 54} (1996) 7705


\bibitem{kn:meyerstm96} G. Meyer, Rev. Sci. Instrum. {\bf 67} (1996) 2960

\bibitem{kn:binnig82} G. Binnig, H. Rohrer, Ch. Gerber and E Weibel, Phys. Rev. Lett. {\bf 49}
(1982) 57

\bibitem{kn:lang88} N. D. Lang, Phys. Rev. B {\bf 37} (1988) 10395

\bibitem{kn:olesen96} L. Olesen, M. Brandbyge, M. R. S{\o}rensen, K W Jacobsen, E.
L{\ae}gsgaard, I. Stensgaard and F. Besenbacher, Phys. Rev. Lett. {\bf 76} (1996)
1485

\bibitem{kn:bennewitz} R. Bennewitz, M. Bemmaerlin, M. Guggisberg, C. Loppacher, A. Baratoff,
E. Meyer, H.-J. G\protect{\"u}nterrodt, Surf. Inerface Anal. {\bf 27} (1999) 462

\bibitem{kn:kresse} G. Kresse and D. Joubert, Phys. Rev. B {\bf 59} (1999) 1758

\bibitem{kn:blochl} P. E. Bl{\"o}chl, Phys. Rev. B {\bf 50} (1994) 17953

\bibitem{kn:kresse_VASP} G. Kresse and J. Furthm\protect{\"u}ller, Phys. Rev. B {\bf 54}
(1996) 11169

\bibitem{kn:pw91} J. P. Perdew, J. A. Chevary, S. H. Vosko, K. A. Jackson, M. R.
Pederson, D. J. Singh and C. Fiolhais, Phys. Rev. B {\bf 46} (1992) 6671


\bibitem{kn:olsson} F. E. Olsson, N. Lorente, M. Persson, L. J. Lauhon and W. Ho, J. Phys. Chem
B, {\bf 106} (2002) 8161

\bibitem{kn:langdIdV} N. D. Lang, Phys. Rev. B {\bf 34} (1986) 5947


\bibitem{kn:chen_prb} C. J. Chen, Phys. Rev. B {\bf 42} (1990) 8841

\bibitem{kn:hofer_prl} W. A. Hofer, A. J. Fisher, R. A. Wolkow and P. Gr\protect{\"u}tter,
Phys. Rev. Lett. {\bf 87} (2001) 236104

\bibitem{kn:desjonqueres} M. C. Desjonqu\protect{\'e}res and D. Spanjaard, Concepts in surface
physics, Springer Verlag, Berlin (1996)

\bibitem{kn:poole} R. T. Poole, J. G. Jenkin, J. Liesegang and R. C. G. Leckey, Phys. Rev. B
{\bf 11} (1975) 5179


\bibitem{kn:note1} Note that a proper notation of the
overlayer with respect to the Cu(311) lattice can be phrased in matrix
notation as (0 3; 1 1) for both structures~\cite{kn:desjonqueres}

\bibitem{kn:note2}The slab contained six layers of Cu atoms.  We used
the same slab for the mono- and bi-layers, where a vacuum region
corresponding to 17 \AA\ for the bare Cu slab was used. For the
trilayer, the vacuum region corresponded to 25 \AA\ for the bare Cu
slab. The surface unit cell contained three Cu atoms in each substrate
layer and two Na and Cl atoms in each NaCl layer. The surface
Brillouin zone (SBZ) was sampled using 36 $k$-points and the
plane-wave cut-off energy was 300 eV.

\bibitem{kn:note3}The atomic positions
of the Na and Cl atoms and the first two layers of the Cu slab were
allowed to relax while the remaining four Cu layers were fixed at
their bulk equilibrium positions

\bibitem{kn:note4}Experimentally measured band
gap for bulk NaCl is 8.5 eV~\cite{kn:poole}.

\bibitem{kn:note5}Ultra violet photoemission and electron energy loss spectroscopy measurements by
S. F\protect{\"o}lsch (PhD thesis) showed a fully developed band gap of about 8.5eV for a bilayer
on NaCl on Ge(100)

\end{thebibliography}
\end{document}